# Evolution of the distance scale factor and the Hubble parameter in the light of Planck's results

Guibert U. Crevecoeur

## 1. Introduction

The classical approach of the evolution of the Universe started with the discovery of its expansion following Hubble measurements in the years 1920. It is based on Einstein's field equations [1] as usually reduced using Robertson – Walker's metrics. As measurements and observations accumulated, the classical model of an hot big bang followed by an evolution in roughly three stages developed (hot big bang flat Universe scenario) : a first stage of inflation (around $10^{-35}$ to $10^{-32}$ s), a second radiation-dominated stage up to the emission of the cosmic microwave background (CMB) radiation (maximum around 380,000 years) and a third matter-dominated stage since. The inflationary stage is more and more assumed because it allows to explain observational data like the flatness of the Universe which are otherwise difficult to explain. During these stages, the temperature decreased down to the present temperature. The equations of state, relating the pressure $p$ to the density $\rho$, are given by $p = -\rho$ for the inflationary stage, $p = (1/3).\rho$ for the radiation-dominated stage and $p = 0$ for the matter-dominated stage. The Hubble parameter $H$ had a value of the order of *$10^{49}$ km/s.MPc* during the inflationary stage, decreasing down to *700,000 km/s.MPc* during the radiation-dominated stage and to a present value around *50 - 100 km/s.MPc* during the matter-dominated stage [2,3,4]. The cosmological constant $\Lambda$ initially introduced by Einstein in his equations to allow a non-evolving Universe was then usually deleted in front of the evidence of the expansion. As the Universe is flat, the spatial curvature factor $k$ in Einstein's equations is put to zero.

From the end of the years 1990, an unexpected discovery made adaptations to the classical model necessary. The acceleration of the expansion of the Universe has first been put into evidence by far $I_a$ supernovae redshift measurements [5–8], then confirmed by the WMAP [9] and Planck [10] missions. This also coincided with increasing frustration about the classical model because the calculated density of the Universe on base of visible baryonic matter was an order of magnitude lower than the critical density. In addition, new observations showed the possible presence of big amounts of invisible matter (called « dark matter ») in galaxies. Therefore, the classical model was adapted in the beginning of the years 2000 by interpreting the acceleration of the expansion as due to « dark energy » and reintroducing the cosmological constant as a possible cause of it. Adding the densities of visible baryonic matter, dark matter and dark energy allows to reach the level of the critical density.

Independently from the classical approach, some analyses were performed assuming the Hubble parameter to be made up of two terms (still in an hot big bang flat Universe scenario) : one term $\beta/t$ constant in space but dependent of time – as is known during the radiation- and matter-dominated stages - and another one $\alpha$ constant in both space and time – as in an inflationary stage [2,3,4]. The Hubble parameter is thus to be read :

$$H = \alpha + \frac{\beta}{t} \qquad (1)$$

It was shown that this « alternative » approach allowed to infer the evolution of the scale factor from an early inflationary stage up to now and in the future. This makes sense as integrating (1) results in a distance scale factor $R \sim e^{\alpha t} t^{\beta}$ giving $R \sim e^{\alpha t}$ (typical for an inflationary stage) for $\beta=0$ and $R \sim t^{\beta}$ (with $\beta = 1/2 \ or \ 2/3$ for the radiation- and matter-dominated stages respectively) for $\alpha=0$. Provided that simple assumptions were made for the transitions between stages, the whole evolution curve could then be drawn. In addition, the values of the main cosmological parameters (deceleration factor, pressure / density ratio) remained during the three stages at values as expected in the classical model existing before the years 2000.

In the classical approach, the evolution of the Universe can be computed back to the beginning. However, no fine tuning of the first stages of inflation and radiation dominated era up to 380,000 hours (CMB) can be performed. As noticed by Mukhanov, the exact de Sitter solution ($H = \alpha = constant$) « fails to satisfy all necessary conditions for successful inflation : namely, it does not possess a smooth graceful exit into the Friedmann stage. Therefore, in realistic inflationary models, it can be utilized only as a zero order approximation. To have a graceful exit from inflation we must allow the Hubble parameter to vary in time. » [11]

Also, Friedmann's solutions for the radiation-dominated era ($H = 1/2t$) and for the matter-dominated era ($H = 2/3t$) – the last known as the Einstein-de Sitter model – do not possess a smooth graceful departure backwards from the inflationary period.

This is one of the reasons why the alternative approach given by Eq. (1) was also examined in parallel. It allows a « graceful exit from inflation » with the Hubble parameter allowed to vary in time while having been constant during the inflationary stage ($\beta/t$ negligibly small towards $\alpha = constant$ during this stage).

Two main hypotheses were then made in the alternative approach as the acceleration of the expansion was not integrated yet in the classical model that time : (a) that there was a drastical pressure drop (factor $10^{-9}$) at the end of the radiation-dominated epoch (atom formation at $T \sim 3000 K$) and (b) that the pressure was strictly zero at present time. One result was that the term $\alpha$ became very small compared to $H$ during the matter-dominated era.

Now that the expansion acceleration has been confirmed by several independent observations, both assumptions are no longer necessary.

Here we compare the results of both approaches in the light of the most recent Planck results [10].

## 2. Method

Because these most recent measurements confirm it, we stay with the hot big bang flat Universe scenario as in previous papers [2,3].

## 2.1. Classical handling (« class » in the figures)

The Lambda-CDM model is used (corresponding to an hot big bang flat Universe - $\Omega_k = 0$) :

The density parameters $\Omega_m$, $\Omega_r$, $\Omega_\Lambda$ and the present Hubble parameter $H_0$ are taken from the 2015 results of the Planck mission [10] :

$\Omega_t$ : Total density parameter (-) ($\Omega_t = \Omega_m + \Omega_r + \Omega_\Lambda = 1$)

$\Omega_m$ : Density parameter for matter (dark + baryonic) ($\Omega_m = 0.308$)

$\Omega_r$ : Density parameter for radiation ($\Omega_r = 0$)

$\Omega_\Lambda$ : Density parameter for dark energy ($\Omega_\Lambda = 0.692$)

$H_0$ : Present Hubble parameter ($H_0 = 67.8\ km/s.MPc$)

### 2.1.1. Computation of the distance scale factor $R$ :

An increment/decrement $dR$ corresponding to an increment/ decrement $dt$ of time is given by :

$$dR = H. \sqrt{\left(\frac{\Omega_m}{R} + \frac{\Omega_r}{R^2} + \Omega_\Lambda . R^2\right)}. dt \qquad (2)$$

with :

- $\Omega_m$, $\Omega_r$, $\Omega_\Lambda$ having the above mentioned values.
- $H$ (in 1/Gyr), the Hubble parameter varying in time according to :

$$H = \frac{\dot{R}}{R} \qquad (3)$$

The curve for $R$ normalized to $R_0$ (i.e. the present value of $R$ which is unknown) is computed cumulatively using an iterative process with decrements/ increments $dt=0.02\ Gyrs$ from present back to the past and forward to the future starting from $R/R_0=1$ and $H_0 = 67.8\ km/s.MPc$.

This brings us to an initial time for the big bang of $-13.8\ Gyrs$.

**2.1.2.** Computation of the deceleration parameter $q$ :

The deceleration parameter is by definition given by :

$$q = -\frac{\ddot{R}.R}{\dot{R}^2} \qquad (4)$$

**2.2. Alternative handling (« alter » in the figures)**

The model with a constant in time term contained in the Hubble parameter is used (for an hot big bang flat Universe) :

The present Hubble parameter $H_0$ and the present time $t_0$ after the big bang are taken from the 2015 results of the Planck mission [10] :

$H_0$ : Present Hubble parameter (***$H_0$ = 67.8 km/s.MPc***)

$t_0$ : Present time from the big bang (***$t_0$ = 13.8 Gyrs***)

**2.2.1.** In order to perform the comparison with the classical results, we consider, in first approximation, that the Universe is matter-dominated back to the CMB (380,000 years after the big bang) with $\boldsymbol{\beta = 2/3}$.

a) Computation of $\alpha$, the term constant in time of the Hubble parameter $H$ :

The value of $\alpha$ is found from (1) at present time $t_0$ :

$$\alpha = H_0 - \frac{\beta}{t_0} = 20.56 \frac{km}{MPc}/s \qquad (5)$$

b) Computation of the distance scale factor $R$ :

The curve for $R/R_0$ is computed cumulatively using an iterative process with decrements/increments $dt=0.02$ Gyrs from present back to the past and forward to the future starting from $R/R_0=1$ using (1) as integrated :

$$\frac{R}{R_0} = e^{\alpha.(t-t_0)} \cdot \left(\frac{t}{t_0}\right)^{\beta} \qquad (6)$$

c) Computation of the Hubble parameter $H$ and of the deceleration parameter $q$ :

The computations are made using (1) for $H$ and (4) combined with (1) and (3) for $q$ to obtain :

$$q = \frac{\beta}{H^2.t^2} - 1 \qquad (7)$$

**2.2.2.** Then the Universe is considered to be radiation-dominated back to the end of the inflationary epoch (assimilated to the big bang, i.e. at − 13.8 Gyrs) with $\boldsymbol{\beta = 1/2}$ and $\boldsymbol{\alpha_{CMB}}$ as deduced from (1) with the value of $\boldsymbol{H_{CMB}}$ reached back to 380,000 years ($\alpha = 428{,}883\ km/s.MPc$). The computation of the other parameters is pursued as above using (6) and (7).

**2.2.3.** Computation of the ratio between the critical densities :

The computation of the ratio between the critical densities in the alternative and in the classical models is made with the formula for the critical density :

$$8.\pi.G.\rho_{crit} = 3.H^2 \qquad (8)$$

With *H* given by (3) in the classical model and by (1) in the alternative model.

## 3. Results

The results of the computations are shown in the following figures (where « class » is for « classical model » and « alter » for « alternative model »).

Figure 1 shows the evolution of the distance scale factor *R/R₀* from the big bang until now and then in the future. It can be seen that the distance scale factor is evolving in a similar way in both models in the past. However, the acceleration in the future becomes higher for the classical model after a few Gyrs.

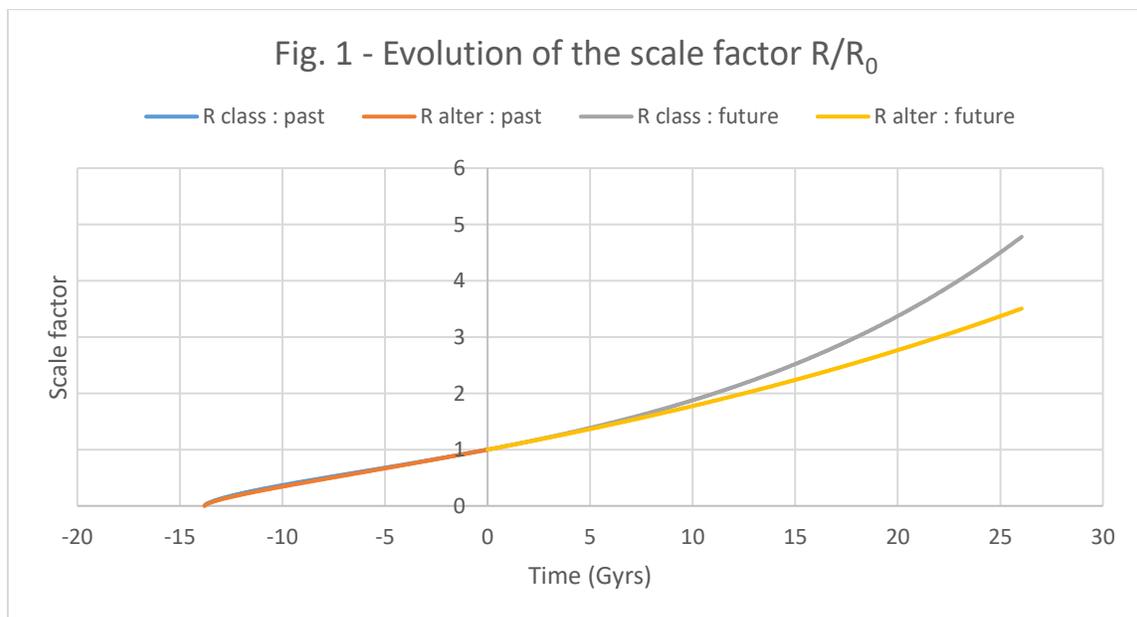

Fig. 1 : Evolution of the distance scale factor *R/R₀* from the big bang until now and in the future

Figure 2 shows the evolution of the deceleration parameter *q*. The acceleration of the expansion starts when *q* becomes negative in the matter-dominated stage. One observes that this takes ($q=0$) place 7.58 Gyrs after the big bang for the classical calculation and a little bit sooner for the alternative model (7.12 Gyrs).

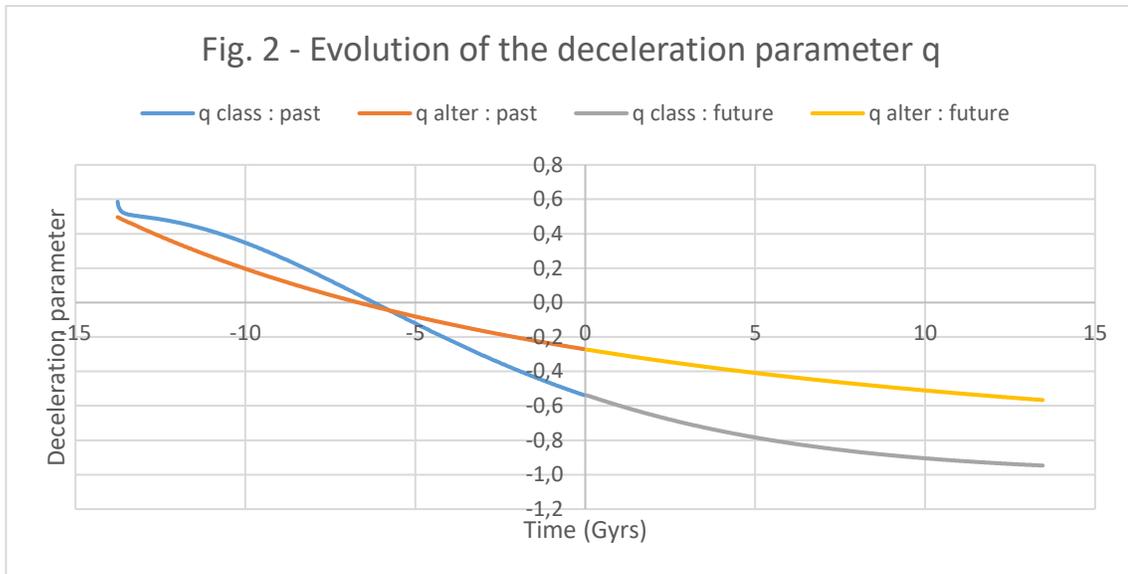

Fig. 2 : Evolution of the deceleration parameter *q*

Figure 3 shows the evolution of the Hubble parameter *H*. The Hubble parameter follows a similar trend for both models. However, starting from the present value of *67.8 km/s.MPc*, it is increasing quicker back into the past and decreasing quicker into the future with the alternative model compared to the classical model. However the difference is small most of the time (7% in the last 10 Gyrs and 10% in the next 5 Gyrs).

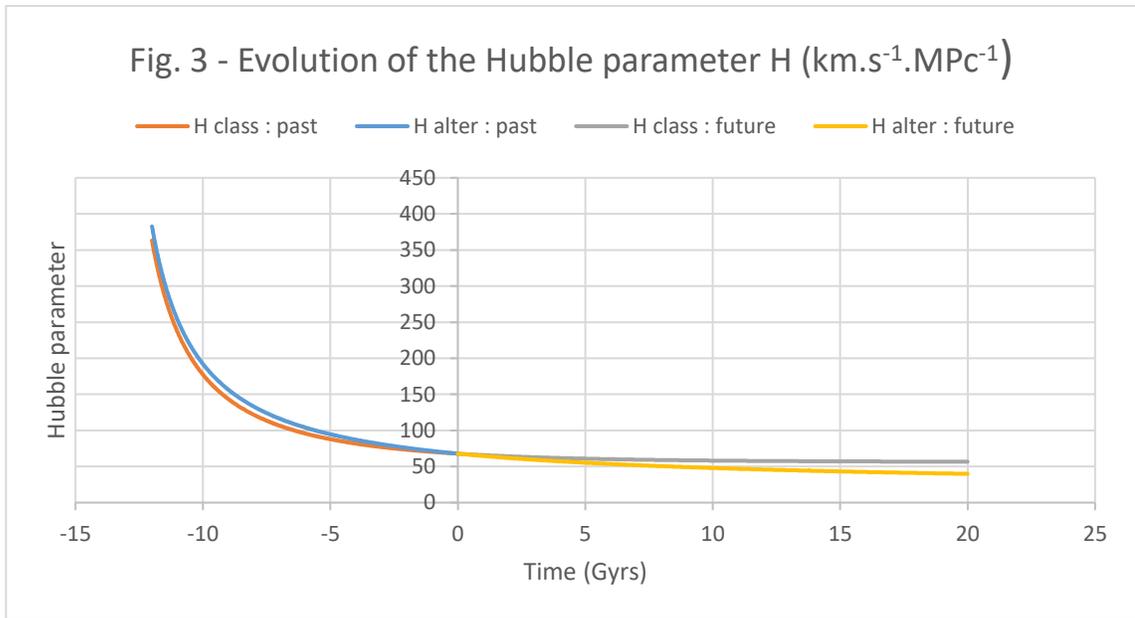

Fig. 3 : Evolution of the Hubble parameter $H$

Figure 4 shows the evolution of the critical densities ratio $\frac{\rho_{crit\ alter}}{\rho_{crit\ class}}$ for the two models as calculated using (8).

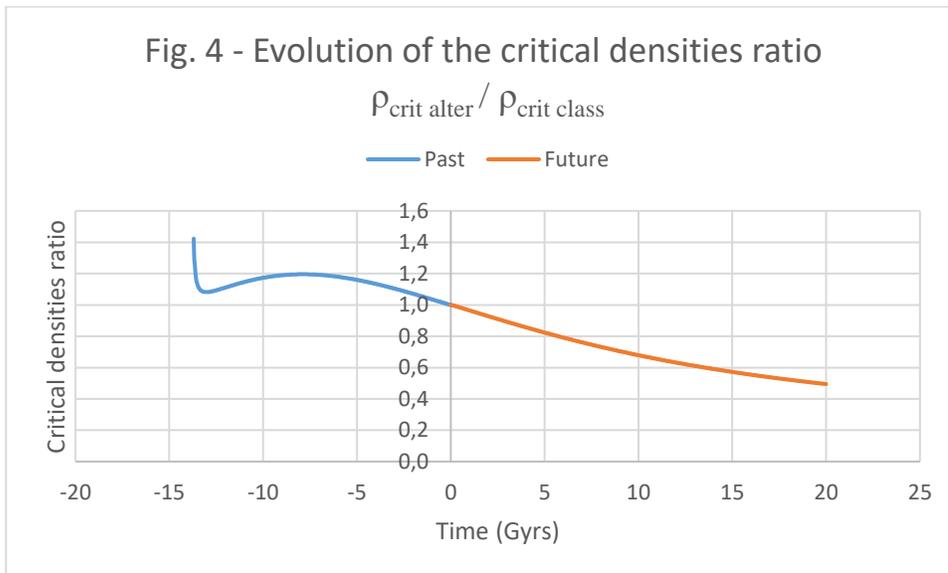

Fig. 4 : Evolution of the critical densities ratio $\frac{\rho_{crit\ alter}}{\rho_{crit\ class}}$

## 4. Comments

We recall that the distance scale factor has been calculated with (2) in the classical approach (where $\Omega_m=0.308$, $\Omega_r=0$ and $\Omega_\Lambda=0.692$ are held constant) and (6) in the alternative approach (where $\alpha=20.56\ km/s.MPc=0.021/Gyr$ and $\beta = 2/3$ are held constant).

In both cases one starts from the present data as measured by Planck's mission : time $t_0=13,800\ Gyrs$ and Hubble parameter $H_0=67.8\ km/s.MPc$. Due to the formula's used, the main difference between both approaches is that there is some refining of the computations in the alternative approach to take account of the specific evolution of $H$ during the radiation-dominated era and of the constancy of $H$ during the inflationary period.

We observe (Fig. 1) that the evolution of the distance scale factor is similar from *-13.8 Gyrs* in the past (big bang) to about *+7 Gyrs* in the future, but that the curves then diverge. The acceleration of the expansion becomes higher in the classical model.

This difference of behaviour is also seen for the deceleration parameter (Fig. 2). This parameter is negative (meaning « acceleration ») for both models during a significant part of the past and in the future, with smaller values for the alternative model showing that the acceleration of the expansion is slower for this model. Both evolution curves intersect at *-5.76 Gyrs* in the past and become positive (indicating « deceleration ») at *-6,22 Gyrs* for the classical approach and *-6.68 Gyrs* for the alternative approach. The effect is then reverse, the deceleration being lower in the alternative model.

The evolution of the Hubble parameter shows a common trend and shape for both models (Fig. 3). However, some discrepancy is found when we check the ratio of the critical density in the alternative model to the critical density in the classical model (Fig. 4). As the critical density is proportional to the square of the Hubble parameter, small differences in this paramer are magnified. If both critical densities (or total densities in a flat Universe) were equal all the time, one would have an horizontal line at ordinate *1.0*. This is clearly not the case. The critical density is higher in the past according to the alternative model and there is even a steep increase approaching the big bang. On the figure, the ratio is of *1.4* at about *-13.7 Gyrs* (which are *100 Myrs* after the big bang, thus still in the matter-dominated stage) but increases drastically towards the past to reach a factor of $10^{26}$ back to a few minutes after the big bang. This is because, in the alternative case, the radiation-dominated epoch could be accounted (from *380 kyrs* back to zero with $\beta = 1/2$ and $\alpha$ constant at a value calculated from the value of *H* at *380 kyrs*). While in the classical simulation, the radiation-dominated stage was not taken into account (for lack of knowledge of the precise values of $\Omega_m$, $\Omega_r$ and $\Omega_\Lambda$ to be used).

However the discrepancy is such that this factor of $10^{26}$ could point to the fact that an inflationary stage would actually have taken place before (with *H* very high of the order of $10^{49}$ [3]).

Now, as $\Omega_m$ and $\Omega_\Lambda$ have been considered constant all the time in the classical computation and there was no requirement of that kind in the alternative case, the slight discrepancy in the past during the matter-dominated era could reflect higher levels of dark matter and dark energy than classically thought. There also seems to be a maximum around *-7 Gyrs*. On the contrary, in the future, things look reversed. A ratio of critical densities lower than *1.0* as shown on the figure would indicate that dark matter and dark energy would lower in the future.

Another interpretation can also be proposed. Normally, according to General Relativity, one would add the energy densities of different matter contributions, in this case one could propose $\rho = \rho_\alpha + \rho_{m,r} = \alpha^2 + \beta^2/t^2$. Up to a factor of $8\pi G/3$, the Hubble parameter would then be given by the square root of the energy density, $H = \sqrt{\alpha^2 + \beta^2/t^2}$. But this is different from Eq. (1). Therefore, in order to be in line with relativistic cosmology using the alternative approach, we used $\rho_{crit\ alt} \approx H^2 = (\alpha + \beta/t)^2$ to build Fig. 4 (up to a factor $8\pi G/3$). We thus assumed that an additional term of the form $2\alpha\beta/t$ is added to the energy densities of different « matter » contributions. This additional term can be interpreted as the contribution of interactions between the remainders of what caused the inflation (given by $\alpha$) and matter/radiation. This could correspond to a form of quantum mechanical interference which would have occurred in the early Universe [12].

Indeed, what occurred in the first instants of the Universe is still matter of speculation. In the classical approach, the quantum mechanical side is assumed to be given by vacuum energy (cosmological constant and density parameter $\Omega_\Lambda$). In the alternative approach, it would be given by a kind of quantum mechanical interference between vacuum energy fluctuations and the different forms of matter and radiation in the first instants of the Universe. This interference then holds for the posterior periods when the Universe has become much larger.

Thus, compared to the classical approach, the cosmological constant can remain nil ($\Lambda = 0$) but a term $\alpha$ constant in time is added in the Hubble parameter $H$ to take account of an inflationary period, as resulting e.g. from a large vacuum energy fluctuation as has been shown by Mukhanov and Chibisov [12]. In 1981, these authors discovered that quantum fluctuations could be responsible for the large scale structure of the universe. They derived the « spectrum of cosmological metric perturbations generated in a de Sitter stage of accelerated expansion (the word "inflation" had not been invented yet at this time) from quantum fluctuations » [13]. The spectrum « came out to be logarithmically dependent on the scale ». Their theoretical prediction was several times experimentally verified since by the slight temperature differences in the Cosmic Microwave Background (CMB) radiation (recently Wilkinson Microwave Anisotropy Probe – WMAP and Planck's satellite missions).

After this inflationary period, the expansion would have gracefully decelerated, but the term $\alpha$ for the contribution of vacuum energy fluctuations would not have fully disappeared. For the purpose of the present computations, we assumed that it would have become constant again at the end of the radiation dominated era. But other scenarii are possible. Because of quantum mechanical interference, this term $\alpha$ would still be entangled with radiation/matter during the relevant periods where radiation and afterwards matter dominated.

This would be similar to the main handling of probability amplitudes as a basic feature in quantum mechanical issues (e.g. two-slits experiments) [14]. Parameter $H$ would reflect probability amplitude for the expansion and include two terms : one (« $\alpha$ ») for vacuum fluctuations including an inflationary episode and another one (« $\beta/t$ ») for either radiation ($\beta = 1/2$) or matter ($\beta = 2/3$) domination. As the followed ways are indistinguishable, there would be interferences between the amplitudes and, thus, the sum of the amplitudes must be squared to get the probability : $|H|^2 = |\alpha + \beta/t|^2$.

As a result, an interference term $\approx 2\alpha\beta/t$ appears reflecting interferences between vacuum fluctuations and either radiation or matter depending on the stage of expansion considered : radiation- or matter-dominated. This gives new insight into « dark energy » which could correspond to such an interaction term contributing to the energy density instead of the cosmological constant. First computations show that this term could amount to 50 % or more of the critical density.

In addition, predictions can be made on the first stages of evolution of the Universe, e.g. that the contribution of « dark energy » would be negligible during the radiation-dominated era and only become increasing again during the matter-dominated era giving an explanation to the « why now » problem (this will be shown in a next article).

## 5. Conclusions

A comparison is made of the evolution of cosmological parameters according to two models in the light of the 2015 results of Planck's mission. The first model is the classical model where density parameters for the matter (baryonic and dark matter) and for dark energy are fixed. The second model is based on the assumption that the Hubble parameter contains a term (mostly) constant in time together with the standard Einstein-de Sitter $2/3t$ term. Adding a term constant in time has the straightforward consequence that the expansion of the Universe is accelerating.

Except around time 0 and after about 7 Gyrs in the future, thus on a period of about 21 Gyrs, there is good coherence of results between both models.

It is however shown that the evolution of the critical density (or the total density in a flat Universe) is different in both models. Because of the differing fixed ingredients in the models, this could point to varying amounts of dark matter and dark energy in the past and in the future, i.e. more in the past and less in the future for the alternative model compared to the classical model.

Another interpretation would be that this discrepancy would reflect a kind of quantum mechanical interference between vacuum energy fluctuations and the different forms of matter and radiation in the first instants of the Universe including an inflationary stage as assumed in the alternative approach. This interference would then hold for the posterior periods when the Universe became much larger and galaxies were developing. It would amount to 50% or more of the critical density.

## Acknowledgments :

I wish to thank Professor Laurent Zimmermann of the Free University of Brussels (ULB) for his example of computation using the classical model (Cours Public d'Astronomie : Cosmologie).